\documentclass{elsarticle}

\usepackage{hyperref}

\usepackage{color}

\journal{Journal of Nuclear Materials}

\begin{document}

\begin{frontmatter}

\title{Influence of kinetic effects on terminal solid solubility of hydrogen in zirconium alloys}

\author{Peter Kaufholz$^{\ast}$, Maik Stuke, Felix Boldt, Marc P\'eridis}
\cortext[mycorrespondingauthor]{Corresponding author}
\ead{Peter.Kaufholz@grs.de, Maik.Stuke@grs.de, Felix.Boldt@grs.de, Marc.Peridis@grs.de}
\address{GRS gGmbH, Forschungszentrum, Boltzmannstr.14, 85748 Garching, Germany}




\begin{abstract}
The integrity of irradiated zirconium based nuclear fuel cladding is related to the precipitation of hydrides which is closely connected with the solubility of hydrogen. A review on the development of measurement technologies is given and the resulting terminal solid solubilities are reflected with respect to the established model of hydrogen solubility in zirconium alloys. The results often allow for a different interpretation than the established model. An alternative qualitative approach is proposed in which the fixed TSSp for the precipitation of zirconium hydrides is replaced by a kinetic model based on thermal history, total hydrogen content and the cooling rate. The influence of the modification of the model from a fixed TSSp to a kinetically limited TSSd is discussed with respect to the understanding of hydride embrittlement processes in long term storage.
\end{abstract}

\begin{keyword}
Hydrides\sep Zirconium \sep Terminal Solid Solubility \sep Spent Nuclear Fuel \sep Embrittlement
\end{keyword}

\end{frontmatter}


\section{Introduction}

A major prerequisite for the dry cask storage of spent nuclear fuel (SNF) is the exclusion of systematic cladding failure. The latter potentially influences storage, transport and necessary steps for the subsequent transfer of the SNF to a final disposal facility. However, during several years of operation of the fuel in a nuclear reactor the fuel cladding material oxidizes and collects hydrogen generated by this oxidation reaction which has a strong influence also on its long term behavior \cite{MOT15}. During its continuous development, zirconium alloys used for light water reactor (LWR) fuel cladding were optimized to reduce waterside corrosion and hydrogen uptake. The continuous development in cladding material allows for higher burn up levels and therefore a better fuel utilization. Nevertheless, the waterside oxide thickness and hydrogen content of LWR fuel cladding fabricated from modern materials (e.g. Zirlo, M5) is low compared to early standard “prime candidate alloy” material as Zircaloy 4 \cite{MOT15, WEI07, MOT16}. The hydrogen introduced into the zirconium alloy results in a solid solution of free hydrogen species in the $\alpha$-Zr metal matrix. The dissolved hydrogen is affected by diffusion processes and remains mobile. At some point the amount of hydrogen introduced into the zirconium alloy exceeds the terminal solubility limit, leading to the formation of zirconium hydrides. The structure of reactor-formed zirconium hydrides depicts a platelet form, precipitating preferably in circumferential direction under in-core pressure and temperature conditions. The equilibrium between free hydrogen and zirconium hydrides is temperature dependent. However, the precipitation of zirconium hydrides in the $\alpha$-Zr matrix can influence the cladding integrity limits, through a process known as hydride “embrittlement”. For a general overview see e.g. \cite{MOT15}. 

Several studies were carried out regarding the behavior of zirconium hydride precipitation and dissolution within reactor transient and accident conditions and the accompanied hydride embrittlement \cite{HER14}. Circumferential zirconium hydrides precipitated in the fuel cladding at the end of the fuel assembly's last cycle and remain during the conditions of forced cooling of spent fuel wet storage in spent fuel ponds. 

In order to prepare spent nuclear fuel for dry storage or transport, an in-cask drying process is applied. The residual heat of the spent fuel is used to heat the cask internals and remove residual water by applying reduced atmospheric pressure and helium flooding. The procedure for in-cask drying may be repeated to achieve the transport and storage criteria. Temperatures during the drying process can become comparable to the operational cladding temperatures resulting in a dissolution of zirconium hydrides. The cask is closed under reduced pressure in a helium atmosphere, leaving the cladding temperatures at elevated temperatures. The slow decrease of temperature and accompanied hydride precipitation in the cladding during dry storage is coupled to the decay heat of the fuel inside. Additionally, the pressure gradient during operation is reversed into an pressure gradient of high rod internal pressure in combination with the cask's low internal pressure.

Both the solvus of zirconium hydrides and the changed orientation of the stress state might cause zirconium hydrides to precipitate in a radial direction (“re-orientation”) during slow cooling in dry storage. The combination of stress state and precipitation behavior negatively influences the cladding integrity due to embrittlement and is the driving force for the delayed hydrogen cracking (DHC) phenomenon \cite{IAE10}. 

The solubility phenomenon can be described as equilibrium between the solid zirconium hydride and the dissolved hydrogen in the matrix. The idealized terminal solid solubility system would be only depending on the temperature of the system. Within the last decades a system was established using two different terminal solid solubility relations for the description of the solubility phenomena of hydrogen in zirconium metals, for increasing and decreasing temperatures, respectively. 

In the following section we review the common methodologies towards prediction and quantification of delayed hydrogen cracking. We discuss the results in section 3 and propose ideas for an alternative model for the reorientation of zirconium hydrides. Section 4 concludes our work and gives an outlook on further research activities and results.

\section{Methodology in zirconium hydride analysis}

\begin{figure}[h]
	\centering
		\includegraphics[width=0.80\textwidth]{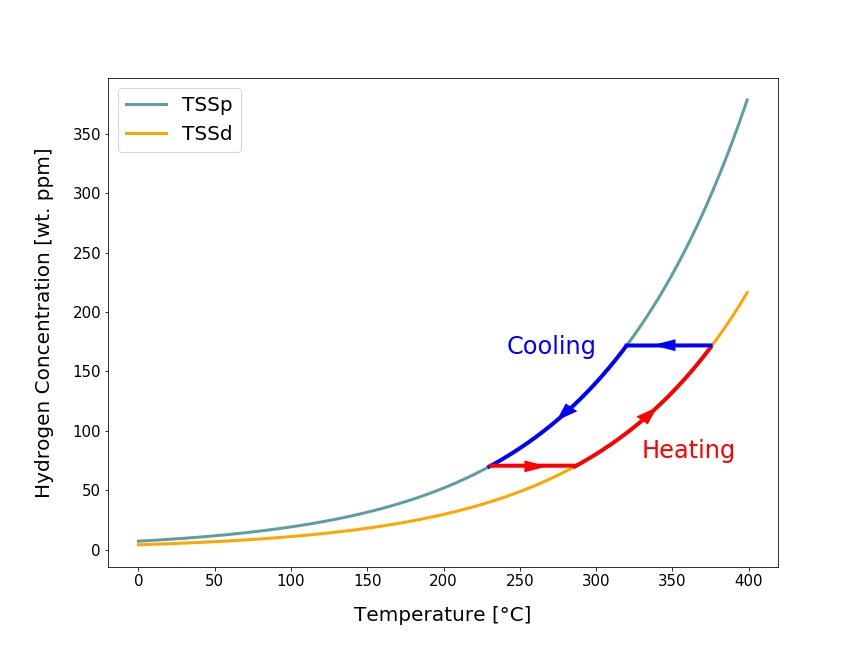}
	\caption{Generic terminal solid solubilities of hydrogen in $\alpha$-Zr-alloys for dissolution (TSSd) and precipitation (TSSp) as a function of the temperature. The course of the hydrogen solubility during a successive heating and cooling process is described in red or blue, respectively. }
	\label{fig:TSSdTSSp}
\end{figure}

The dissolution terminal solid solubility (TSSd) describes the solubility of hydrogen in Zr-alloys during the heating process. Additionally, the precipitation terminal solid solubility (TSSp) describes the precipitation of zirconium hydrides in Zr-alloys observed with decreasing temperature during the cooling process \cite{MAR71}. This effect of different dissolution and precipitation solubilities observed with thermal cycling was called hysteresis. Figure \ref{fig:TSSdTSSp} shows a generic plot of TSSd and TSSp and the progress followed by a specimen during heating and successive cooling.

Focusing on the thermodynamic equilibrium of the solubility of hydrogen, the hysteresis arises because additional energy is necessary in the precipitation process, compared to the dissolution process. This energy was believed to be used by the Zr metal matrix, which needs to be expanded for the crystallization of the zirconium hydride. Additionally, surface energy is discussed to contribute to the effect of hysteresis \cite{BAR13}. Puls et al. \cite{PUL12} suggested this accommodation energy splits into parts for the elastic and plastic deformation of the surrounding Zr lattice. Hydride nucleation begins in a stress-free state and deforms the Zr lattice elastically. Further growth of the hydride requires plastic deformation of the metal lattice, which cannot be recovered in the following dissolution cycle.

\subsection{Diffusion Techniques}

Early experiments focusing on the determination of the effect of temperature on the terminal solid solubility of hydrogen in alloys were carried out either as diffusion couple experiments or as thermo-diffusion experiments \cite{KEA67, SAW67, SAW60}. The diffusion couple experiments use the thermal diffusion of dissolved hydrogen between a hydrogen charged and an uncharged specimen connected by welding. The hydrogen-rich metal contains a higher hydrogen concentration than the expected terminal solid solubility at the temperature of the experiment. The welded specimens were afterwards heated to allow for hydrogen to diffuse into the uncharged specimen of the couple. As diffusion only takes place with dissolved hydrogen, one can calculate the terminal solid solubility of hydrogen at the diffusion temperature knowing the amount of hydrogen diffused into the hydrogen free specimen. From the hysteresis point of view, the diffusion couple is heated to dissolve zirconium hydrides at the level of TSSd into the matrix. After reaching diffusion equilibrium, a cool-down of the couple would lead to hydrogen concentrations equal to the TSSd in the low hydrogen part of the specimen. In thermo-diffusion experiments, a hydrided specimen is exposed to a temperature gradient where the distribution of hydrogen is guided by thermo-diffusion. At a certain point within the gradient, the solubility of hydrogen is exceeded, resulting in the precipitation of zirconium hydrides. The hydrogen concentration at the transition between the area containing zirconium hydrides and dissolved zirconium area is taken as terminal solid solubility. However, from the hysteresis point of view, the terminal solid solubility derived by this method at the boundary concentration to precipitation depicts the TSSp. 

Before TSSd and TSSp were fully defined, Kearns \cite{KEA67} discussed the observation of an apparent supersaturation area describing the region of higher dissolved hydrogen concentrations above the bulk measurements. Additionally, \cite{KEA67} compared their data depicting TSSd to literature data including the data of \cite{SAW60}, showing slightly increased values for the data depicting TSSp. The resulting limit of supersaturation in \cite{KEA67} can be seen as an early definition of the TSSp. However, measurement uncertainties for the data were quite high, barely allowing the formulation of a defined supersaturation line. It is important to note, that \cite{KEA67} prevented the supersaturation of their high hydrogen content specimen by applying slow cooling rates. The absence of supersaturation in slowly cooled hydride specimen gave a first idea of a kinetic hindrance affecting the terminal solid solubility of hydrogen and leading to the definition of the TSSp.

\subsection{Dilatometry}
Another method for the determination of hydrogen solubility in zirconium alloys is dilatometry. Herein, the change in dimensions of a specimen during hydride precipitation is measured. As zirconium hydrides have a significantly lower density compared to the $\alpha$-Zr matrix in the alloy, every precipitation of zirconium hydrides therefore results in a deformation of the lattice yielding in an overall change in length of the specimen. In dilatometric hydride analysis, hydrided specimen are heated while the change in length is recorded as a function of the temperature. Measurements can be carried out during heating and cooling, depicting the TSSd in the heating experiment mode and TSSp in the cooling experiment mode. Due to the direct response in the measurement and the possibility to perform consecutive heating and cooling experiments using the same material, the method is ideal for the determination of the TSS-curves. The work presented in \cite{SLA67} found a clear hysteresis for different zirconium alloys using dilatometry. They additionally stated that there is no significant dependence of the TSSp to the cooling rate within the range of cooling rates applied in their experiments. However, the method is sensitive to deviations in the type of zirconium hydrides as they exhibit different densities. The stoichiometry of zirconium hydrides was found to be affected by the cooling rate \cite{TUL12}.

\subsection{Differential Scanning Calorimetry}

With the further development of measurement technologies, new methods became available to analyze the solubility of hydrogen in zirconium alloys. The technology of differential scanning calorimetry (DSC) opened a new option for the determination of the hydrogen terminal solid solubility. The method is based on calorimetric measurements of absorbed heat during a heating process. The detection of the affiliated heat of the specimen allows for the quantitative description of the dissolution of hydrides. However, the method is sensitive to deviations in the type of zirconium hydrides as they exhibit different formation enthalpies. An additional factor of uncertainty in this technique is the interpretation of the measurements raw data. As shown in \cite{VIZ02} a variation of about 20 $^\circ$C in TSS-values can be related to different interpretation. An often cited article in the field of hydrogen solubility is \cite{MCM00} who performed extensive work on the solubility of hydrogen with respect to welded zirconium alloys using DSC. McMinn et al. \cite{MCM00} found the TSSp to be sensitive to the holding time and maximum temperature in their DSC-experiments. A variation of the cooling rate in the range between 0.5 $^\circ$C/min and 10 $^\circ$C/min does only show little effect on the TSSp-curve. Using DSC-method, \cite{TAN} found that the precipitation rate of hydrides at TSSp is controlled by the diffusion of hydrogen.

\subsection{X-ray Methods}
The modern application of X-ray diffraction for the analysis of hydride re-orientation and hydride precipitation opens an option to get a direct response from zirconium hydrides. Using X-ray diffraction it is possible to observe hydride dissolution and precipitation kinetics in-situ at high temperatures. The method is extensively used to investigate the effect of stress on the crystallization orientation of zirconium hydrides \cite{CIN17, BLA15, DAU09}. 
Colas et al. \cite{COL10} applied in-situ X-ray diffraction on the dissolution and precipitation of zirconium hydrides and compared it to DSC-measurements. However, the cooling rates were faster compared to the DSC cooling rates. Blackmur et al. \cite{BLA15} also applied in-situ analysis on the precipitation of zirconium hydrides. They found TSSp-values for fast cooling rates to be different from the literature \cite{MCM00}. A subsequent hold time after the fast cooling then leads to TSSp-values consistent with the literature. 

While the determination of TSSp in the traditional methods is linked to a full dissolution of zirconium hydrides, followed by a cooling process to obtain the first precipitation, in-situ X-ray analysis opens an option to observe the precipitation within a specimen with existing zirconium hydrides. Colas et al. \cite{COL14} observed the dissolution and precipitation of zirconium hydrides with in-situ X-ray diffraction and found generally good agreements with DSC methods. Nevertheless they found the hysteresis to disappear if cooling takes place with zirconium hydrides present \cite{COL14}.

\section{Summary and Discussion}
The terminal solid solubility of hydrogen in zirconium alloys is described as a hysteresis model exhibiting two terminal solubilities for dissolution and precipitation, respectively. However, the constant development of measurement technologies allowed for more precise measurements. The re-interpretation of the available data in combination with a consideration of the uncertainties allows for some discrepancy of the data from the established model of the hysteresis developed in the 1970’s \cite{MAR71}. Therein, the fixed TSSp with a hysteresis was explained by an energy deficiency necessary to overcome the hydride matrix volumetric misfit strain. Starting with the TSSd depicting a limit being most probably defined by thermodynamics, the TSSp shows far more deviation in the measured data. Pan et al. \cite{PAN} suggested a splitting of the TSSp into a TSSp1 and TSSp2 wether there are zirconium hydrides available or not. The results of \cite{COL14} support a preferred precipitation of zirconium hydrides on existing hydride crystals. The hydride tip introduces a tensile strain in the nearby lattice that lowers the accommodation energy barrier necessary for further hydride to precipitate there. The growth of existing hydrides is therefore preferred over the nucleation of new hydrides, the latter having to overcome the matrix-hydride volumetric misfit strain. 

\begin{figure}[hp]
	\centering
		\includegraphics[width=1.00\textwidth]{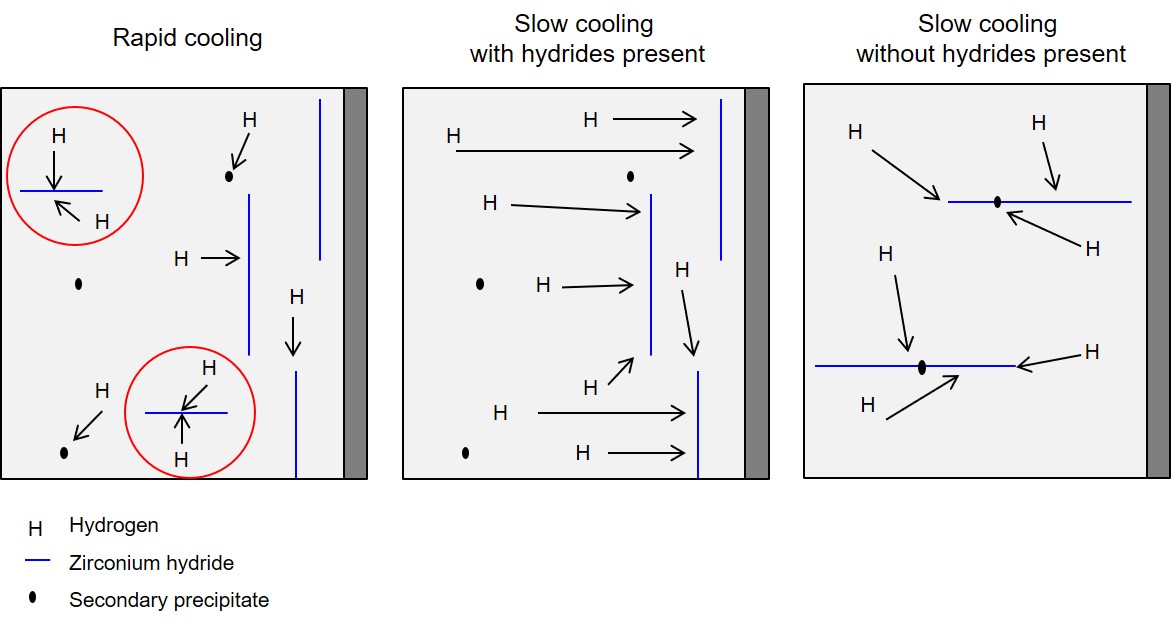}
	\caption{Sketch of the three different scenarios as a result of the kinetic effect in combination with different boundary conditions. Light grey depicts metallic $\alpha$-Zr and dark grey the outer cladding’s surface oxide layer (ZrO$_2$). The re-orientation of zirconium hydrides is related to the Zr-matrix strain. Left: Rapid cooling (e.g. experimental rates 1-10$^\circ$C/min) of the specimen leads to hydride growth in combination with spontaneous precipitation developing a comparatively high hysteresis (red circles). Middle: Slow cooling (e.g. during fuel storage with rates of 1-20$^\circ$C/a) of the specimen with hydrides present may allow hydrogen to diffuse to existing hydrides resulting in their growth with a comparatively low hysteresis. Right: Slow cooling after full dissolution of hydrogen will lead to the nucleation of few hydrides with minimal hysteresis while dissolved hydrogen is able to migrate and support hydride growth.}
	\label{fig:sketch}
\end{figure}

An important factor for the description of hydride precipitation is the diffusion kinetics of dissolved hydrogen to allow dissolved hydrogen to migrate to an existing hydride structure to precipitate. As a result of the data review, TSSp may better be described as a deviation of the thermodynamic TSSd influenced by kinetics, enhanced by the strain-energy variations depending on microstructural state. The kinetic hindrance resulting in the TSSp may be linked to the diffusion kinetics of dissolved hydrogen in the matrix. The results of \cite{MCM00} show an increase of the TSSp with increasing peak temperature and holding time, supporting the finding of hydrogen diffusion playing a role in the kinetic effect. However, this cannot be taken as a proof because the annealing of structure damages, which reduce the matrix-hydride volumetric misfit, is also related to time and temperature. Blackmur et al. \cite{BLA15} showed, that a fast quench of temperature would lead to significantly higher TSSp-values compared to those known from the literature. However, holding at an elevated temperature after a partial quench leads again to the literature data, which can be taken as proof for a diffusion guided kinetic effect. Experiments described elsewhere in the literature were carried out deploying cooling rates between 1 $^\circ$C/min and 10 $^\circ$C/min, which are several orders of magnitude higher compared to the cooling rates in dry storage of spent nuclear fuel. Barrow et al. found a kinetic limitation for the TSSp at high temperatures and for fast cooling rates \cite{BAR13}. Combining the theory of TSSp being affected by kinetics with the discrepancy in cooling rates between experiments and storage asks for the development of a kinetics based model to predict hydrogen solubility behavior in long term storage. Figure \ref{fig:sketch} shows a qualitative overview of the possible consequences of a kinetic effect on the precipitation of hydrides divided in three possible scenarios.

Applying a rapid cooling in the left hand scenario of figure \ref{fig:sketch}, several small super-saturated areas are generated due to the limited time for dissolved hydrogen to migrate. Within the supersaturation, nucleation of new zirconium hydrides causes a hysteresis due to the energy necessary to overcome the hydride matrix volumetric misfit and the surface energy for the creation of a heterogeneous surface in the $\alpha$-Zr matrix. Applying lower cooling rates in the middle scenario of figure \ref{fig:sketch} will lead to longer migration pathways. With existing hydrides in the matrix, the precipitation is believed to preferentially take place at the existing hydrides with just comparatively low hysteresis caused by the volumetric misfit. Within the last option, slow cooling is applied after full dissolution of hydrides. The slow cooling will lead to the nucleation of few hydride crystals which tend to grow slowly due to the migration of hydrogen out of the matrix. Within this option, the nucleation may be linked to lattice distortions by secondary phases or phase boundaries, leading to only comparatively low hysteresis caused by the volumetric misfit strain. In the case of an $\alpha$-Zr matrix under tensile stress, this hysteresis may be reduced by the Zr matrix strain. This may even lead to a precipitation of Zr-hydrides without hysteresis, due to the opposite nature of the effects. For this scenario a triaxial tensile stress state would be necessary, which is unlikely to occur in a cladding material used in a fuel rod. The orientation of new nucleations of zirconium hydrides may be strongly related to the Zr-matrix strain, being the reason for re-orientation.

Recent literature \cite{DEN18, LAC18} support a kinetic hindrance being the main reason for the formation of a TSSp as a variation of the TSSd defined by thermodynamics. The effect of strain-energy misfit becomes less influential. However, the anisotropy of the zirconium matrix related to the diffusion of hydrogen and the extent of surface energy for the creation of a heterogeneous surface in the Zr-matrix were not taken into account for the model development.

\section{Conclusion and Outlook}
In concordance with available experimental data we propose the introduction of a kinetic adjustment  in the model for the precipitation of zirconium hydrides in Zr-alloys. The precipitation rate of zirconium hydrides at existing zirconium hydride positions is guided by the diffusion of hydrogen. However, if the cooling rate exceeds a limit which does not allow hydrogen to migrate to the existing hydride formations, local supersaturation takes place. If the chemical potential of the supersaturated hydrogen solution exceeds the volumetric hydride matrix misfit strain, the new nucleation of zirconium hydrides is initiated, causing the hysteresis described by the TSSp. Due to its thermodynamic origin, the TSSd is considered to represent the lowest possible precipitation solubility curve. The solvus for precipitation of hydrogen in zirconium alloys is influenced by the volumetric misfit, surface energy and kinetics to form TSSp. However, material structure, stress state and cooling rates under dry storage conditions have a mitigating effect on the hysteresis, which may lead to a precipitation of zirconium hydride at the level of TSSd for conditions similar to dry storage conditions.

The implementation of a kinetic term in the established model does effect the re-orientation of zirconium hydrides during long term dry storage.  Additionally, the effects caused by material irradiation and material morphology have to be considered as well  to enable for a reliable prediction of the hydride behavior.  The implementation of a kinetic term for the precipitation of zirconium hydrides and the effect on the nucleation of new zirconium hydride crystals may also change the understanding of hydride re-orientation. 

Further statistical analysis of experimental data with respect to measurement reliability and data significance might support the proposed kinetic adjustment. The proposed model may lead to open questions concerning the nucleation, orientation and growth of hydrides at low cooling rates and full dissolution of hydrides. Zirconium hydride nucleation may also be related to the properties of secondary phases and grain boundaries in Zr-alloys. 

From the safety aspect of dry storage of SNF, further investigations of the proposed kinetic model enhancement is needed. Considering for instance modern cladding materials with low hydrogen content the proposed model may lead to the preferential formation of radial hydrides. This in turn might lead to a significant change of the predicted cladding integrity.

\section{Acknowledgment}

The work was partially funded and supported by the German Federal Ministry for Economic Affairs and Energy under grant No.~RS1552 and the German Federal Ministry for the Environment, Nature Conservation and Nuclear Safety under grant No.~4715E03310. 

\section*{References}

\bibliography{TSS_dp_arxiv}

\end{document}